\documentclass[superscriptaddress,secnumarabic,
amssymb,amsmath,nobibnotes,aps,prd,showkeys,showpacs,nofootinbib,twocolumn]{revtex4}%
\usepackage{graphicx}
\usepackage{epsf}
\usepackage{bm}
\usepackage{amsmath}
\usepackage{amsfonts}
\usepackage{amssymb}
\usepackage{epstopdf}
\usepackage{natbib}%
\providecommand{\U}[1]{\protect\rule{.1in}{.1in}}
\newcommand{\be}{\begin{equation}}
\newcommand{\ee}{\end{equation}}

\newcommand{\mincir}{\raise
-3.truept\hbox{\rlap{\hbox{$\sim$}}\raise4.truept\hbox{$<$}\ }}
\newcommand{\magcir}{\raise
-3.truept\hbox{\rlap{\hbox{$\sim$}}\raise4.truept\hbox{$>$}\ }}

\usepackage{color}

\begin{document}

\title{Connecting inflation with late cosmic acceleration by particle production}
\author{Rafael C. Nunes}
\email{rafadcnunes@gmail.com}
\affiliation{CAPES Foundation, Ministry of Education of Brazil, Bras\'ilia - DF 70040-020, Brazil}
% \affiliation{High Energy Physics Group, Departament d'Estructura i Constituents de la Mat\'eria,
% Univ. de Barcelona, Av. Diagonal 647 E-08028 Barcelona, Catalonia, Spain}
\pacs{ }
%%%%%%%%%%%%%%%%%%%%%%%%%%%%%%%%%%%%%%%%%%%%%%%%%%%%%%%%%%%%%%%%%%%%%%%%%%%%%%%%%%%%%%%%%%%%

%%%%%%%%%%%%%%%%%%%%%%%%%%%%%%%%%%%%%%%%%%%%%%%%%%%%%%%%%%%%%%%%%%%%%%%%%%%%%%%%%%%%%%%%%%
\begin{abstract}
\noindent

A continuous process of creation of particles is investigated as a possible connection between the inflationary stage
with late cosmic acceleration. In this model, the inflationary era occurs
due to a continuous and fast process
of creation of relativistic particles, and the recent accelerating phase is driven by the non-relativistic matter creation
from the gravitational field acting on the quantum vacuum, which finally results in an effective equation of state 
less than $-1$.
Thus, explaining recent results in favor of a phantom dynamics without the need of any modifications in the
gravity theory has been proposed. Finally, we confront the model with recent observational data of 
type Ia Supernova, history of the Hubble parameter, baryon acoustic oscillations, and the cosmic microwave background.

\end{abstract}
%%%%%%%%%%%%%%%%%%%%%%%%%%%%%%%%%%%%%%%%%%%%%%%%%%%%%%%%%%%%%%%%%%%%%%%%%%%%%%%%%%%%%%%%%%%%

\maketitle
%%%%%%%%%%%%%%%%%%%%%%%%%%%%%%%%%%%%%%%%%%%%%%%%%%%%%%%%%%%%%%%%%%%%%%%%%%%%%%%%%%%%%%%%%%%%%%
\section{Introduction}
\label{sec:intro}

Recently, Planck Collaboration \cite{Planck} has presented us with the most complete
image of the early Universe. This provides strong constraints
on the inflationary phase. In particular, the spectral index and the tensor-to-scalar ratio have
been measured to be, $n_s = 0.9603 \pm 0.0073$ (68 $\%$ CL), and $r < 0.10$ at a $95 \%$ C.L. respectively
If confirmed, it will lead to important consequences, particularly,
many inflationary models may be ruled out by these bounds.
\\

In the last few years, a large amount of observational data coming from
Type Ia Supernovae (SNe Ia) \cite{Permutter,Reiss11}, Cosmic Microwave Radiation Background (CMB)
\cite{Spergel,ade} and Large Scale Structure (LSS) \cite{Daniel,Percival} reveal that the
universe is currently undergoing through %at
an accelerated expansion.
A crucial quantity in the models based on Einstein gravity
aimed at accounting for this
expansion is the equation of state (EoS) of dark energy
(DE), i.e., $w_{de}= p_{de}/\rho_{de}$, the ratio of the pressure to the energy density
of the dark energy. %the agent responsible for the accelerated expansion.
In the case of the Lambda cold dark matter, $\Lambda$CDM model, this agent can be identified
with the energy of the quantum vacuum whence the corresponding EoS parameter is
just $w_{\Lambda}=-1$. In spite of the observational success of this model, recent
model-independent measurements of $w_{de}$ seem to favor a
slightly more negative EoS (see, e.g., Refs. \cite{Rest,Xia,Cheng,Shafer}), which,
if confirmed, would invalidate the model. In particular, the
Planck mission yields $w_{de} =-1.13^{+0.13}_{-0.10}$ \cite{ade},
Rest et al., using supernovae type Ia (SN Ia) data from
the Pan-STARRS1 Medium Deep Survey found $w_{de} = -1.166 ^{+0.072}_{-0.069}$, i.e.,
$w_{de} < - 1$ at 2.3 $\sigma$ confidence level \cite{Rest}, Shafer and
Huterer \cite{Shafer} using geometrical data
from SN Ia, baryon acoustic oscillations (BAOs), and CMB,
determined $w_{de} < -1$ at 2$\sigma$ confidence level.
The simplest way to get $w_{de} < -1$, consistent with general relativity,
is assume that the DE has your origin at a phantom field \cite{Caldwell}. Despite being compatible with observational data,
such a possibility has serious theoretical problems \cite{Carroll,Cline,Hsu,Sbisa,Dabrowski}.
Other stringent constraints on the EoS have been analyzed in several contexts, such as, considering a variable EoS for both dark components \cite{Kumar},
using cosmographic methods to investigate various dynamic EoS models \cite{Aviles}, and a reconstruction of the dark energy
EoS via node-based reconstruction \cite{Vazquez}.
\\

In a recent work \cite{Dr1}, the authors investigated an alternative to this possibility, namely,
that the measured equation of state, $w_{de}$,
is in reality an effective one, the equation of state of
the quantum vacuum, $w_{\Lambda}= -1$, plus the
negative equation of state, $w_c$, associated to the
production of particles by the gravitational field acting
on the vacuum. From a joint analysis of data Supernova type Ia, gamma ray bursts,
baryon acoustic oscillations, and the Hubble rate,
it was obtained that $w_{eff}(z = 0) = -1.155^{+0.076}_{-0.080}$ at 1$\sigma$.
\\

In this context, %one way wander of the primeval accelerated expansion of the universe (i.e., the inflationary era)
one may wonder that the primeval accelerated expansion of the universe (i.e., the inflationary era) could result in
%resulted from 
a very fast rate of relativistic particle productions due to action of the gravitational field
on the quantum vacuum. If for a sufficiently long period of time, this rate was high enough to compensate, or nearly compensate,
for the dilution of particles due to the universe expansion, then the energy density of the particles fluid wound remain
nearly constant giving rise to an inflationary expansion. To the best of our knowledge, this idea was first proposed
in \cite{Turok}.
\\

It is important to mention that inflation driven by the particle production is not a new subject. 
Particle creation as a source of inflation in the early Universe, was first investigated in \cite{Starobinsky1}, 
using the expressions for the energy-momentum of the created particles and the rate of their creation ($\propto R^2$ 
for non-conformal particles in a FLRW universe) derived earlier in \cite{Starobinsky2}. However, it appeared that such model presents some problems, such as, they can not produce a sufficiently low curvature during inflation, and a graceful exit from it. Thus, success in constructing viable inflationary models %which prediction 
was achieved in the Starobinsky model \cite{Starobinsky3} in which the dissipation and the creation of matter occurred already after the end of inflation. However, the idea of particle creation driving inflation was revived after that under the name of warm inflation \cite{warm_inflation}.
The aim of this paper is to explore a possible connection between the inflationary stage
with late cosmic acceleration through of a continuous process of creation of particles by the gravitational field acting on the
quantum vacuum \cite{ccdm_inflation}.
\\

This paper is organized as follows. The next section briefly sums up the phenomenological basis of particle
creation in an expanding homogeneous and isotropic, spatially flat, universe. In section \ref{sec:inflationary_epoch},
we describe the inflationary era as a result of a continuous and fast process of creation of relativistic particles.
In section \ref{sec:inflationary_epoch}, we present the dynamics of the universe in the post inflationary era in presence
of a continuous matter creation associated with the production of particles by the gravitational field acting on the vacuum.
Section \ref{Statistical_Analysis} specifies the various sets of data and the statistical analysis employed to constrain the model.
The concluding section summarizes and gives comments on our findings. As usual, the scale factor of
the Friedmann-Robertson-Walker (FRW) metric is normalized so that $a_0=1$, the naught subscript indicates the present time.

\section{Cosmological models with particle creation}\label{sec:models}

As investigated  by Parker and collaborators
\cite{Parker}, the material content of the Universe may have had
its origin in the continuous creation of radiation and matter from
the gravitational field of the expanding cosmos acting on the
quantum vacuum, %this 
regardless of the relativistic theory of
gravity assumed. In this picture, the produced particles draw
their mass, momentum and energy from the time-evolving
gravitational background which acts as a ``pump" converting
curvature into particles.
\\

\noindent Prigogine  \cite{Prigogine} studied how to insert the
creation of matter consistently in Einstein's field equations.
This was achieved by introducing in the usual balance equation for
the number density of particles, $ (n\, u^{\alpha})_{; \alpha}=0$,
a source term on the right hand side to account for the production of
particles, namely,
%%%%%%%%%%%%%%%%%%%%%%%%%%%%%%%%%%%%%%%%%%%%%%%%%%%%%%%%%%%%%
\begin{equation}
(n\, u^{\alpha})_{; \alpha} = n \Gamma \, ,
\label{Eq:nbalance}
\end{equation}
%%%%%%%%%%%%%%%%%%%%%%%%%%%%%%%%%%%%%%%%%%%%%%%%%%%%%%%%%%%%%%
where $u^{\alpha}$ is the particle fluid four-velocity normalized so
that $ u^{\alpha}\, u_{\alpha} = 1$, and $\Gamma$ denotes the
particle production rate. According to Parker's theorem, the production of
relativistic particles is strongly suppressed in the radiation era \cite{Parker-Toms}.
The above equation, when combined with the second law of
thermodynamics, naturally leads to the appearance of a negative
pressure, the creation
pressure $p_{c}$, which adds to the other pressures (i.e.,
radiation, baryons, dark matter, and vacuum pressure) in the
stress-energy tensor. These results were subsequently discussed
and generalized in \cite{Lima1}, \cite{mnras-winfried}, and
\cite{Zimdahl} by means of a covariant formalism, and were further
confirmed by using relativistic kinetic theory
\cite{cqg-triginer,Lima-baranov}.
\\

Since the entropy flux vector of matter, $n \sigma
u^{\alpha}$, where $\sigma$ denotes the entropy per particle, must
fulfill the second law of thermodynamics $(n \sigma u^{\alpha})_{;
\alpha} \geq 0$, the constraint $\Gamma \geq 0$ readily follows.

For a homogeneous and isotropic universe, with scale
factor $a$, in which there is an adiabatic process of particle
production \footnote{Originally introduced by I. Prigogine et al., \cite{Prigogine}
and after investigated by several authors, the process is adiabatic because constrains the
formulation in which the specific entropy (per particle) is constant. Thus, if the specific
entropy is a constant $\sigma$, its variation with respect to the cosmic time is null, i.e., $d\sigma/dt = 0$.
This adiabatic matter creation process corresponds to an irreversible energy flow from the
gravitational field to the created matter constituents.} from the quantum vacuum, %there exists 
a direct relationship between the creation pressure
and the particle production rate exists as \cite{Lima1,Zimdahl}
%%%%%%%%%%%%%%%%%%%%%%%%%%%%%%%%%%%%%%%%%%%%%%%%%%%%%%%%%%%%%%%%%%%%%%%%%%%%%%
\begin{equation}
\label{pressure_creation} p_c= - \frac{\rho \, + \, p}{3H} \, \Gamma \, .
\end{equation}
%%%%%%%%%%%%%%%%%%%%%%%%%%%%%%%%%%%%%%%%%%%%%%%%%%%%%%%%%%%%%%%%%%%%%%%%%%%%%%%
Therefore, being $p_{c}$ negative, it may have produced the accelerated expansion in the early Universe 
(i.e., inflationary era),
as well as, it may also drive the present accelerated cosmic expansion. Here, $\rho$,
$p$, respectively, denote the energy density and the pressure of
the corresponding fluid, $H=\dot{a}/a$  is the Hubble factor, and
as usual, an overdot denotes the differentiation with respect to
cosmic time.
\\

The EoS associated with the process of creation of matter follows from Eq. (\ref{pressure_creation})

\begin{equation}
\label{eos_creation}
  w_c = - (1+w) \frac{\Gamma}{3H},
\end{equation}
where $w=0$ for non-relativistic matter, and $w=1/3$ for relativistic matter.

\section{The inflationary epoch}
\label{sec:inflationary_epoch}

We consider a spatially-flat Friedmann-Robertson-Walker (FRW) universe, with
Hubble factor $H(t) = ̇\dot{a}(t)/a(t)$. We assume a universe undergoing a
continuous process of particle creation thanks to the action of the gravitational field on the quantum vacuum.
The first Friedmann equation reads

\begin{equation}
\label{H}
  H^2 = \frac{1}{3 M^2_{pl}} \rho,
\end{equation}
where $M_{pl}= (8 \pi G)^{-1/2}$ is the reduced Planck mass.

Let us consider the possibility that the early inflationary phase was the result of a continuous and 
fast process of creation of particles, so fast that the energy density $\rho$ stays practically 
constant by about 55 e-folds, to decline quickly afterwards around the GUT era.

\begin{figure}
\centering
\includegraphics[width=3.2in, height=2.4in]{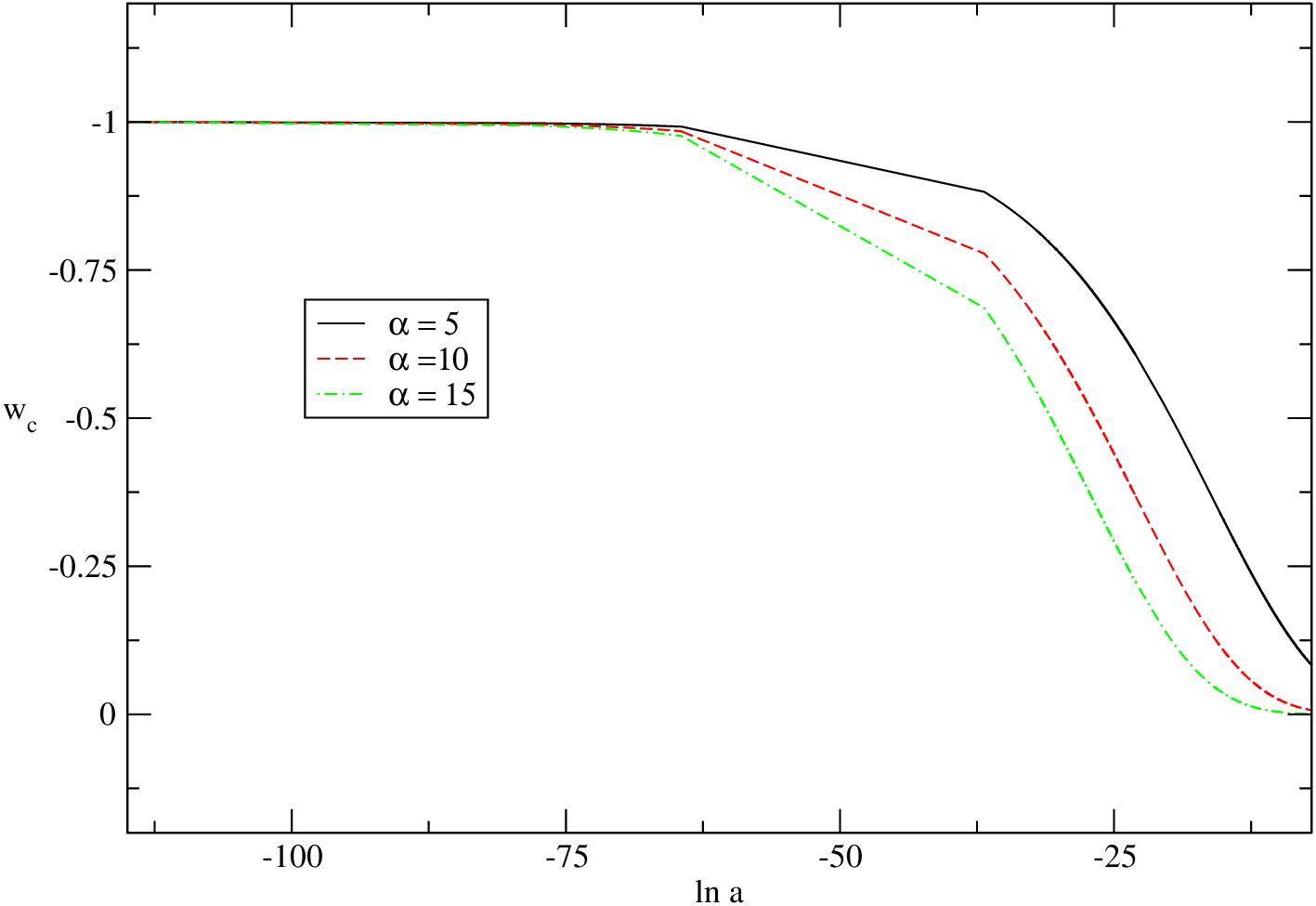}
\caption{\label{Gamma_3H} Evolution of $w_c$ for different values of the parameter $\alpha$.}
\end{figure}

To go ahead, an expression for the rate $\Gamma$ is needed. We first assume that the total rate split into two 
terms as

\begin{equation}
\label{G}
\Gamma= \Gamma_r + \Gamma_{dm},
\end{equation}
with $\Gamma_r$ referring to the particle creation rate in the early universe, and $\Gamma_{dm}$,
the particle creation rate of dark matter particles.
Then we adopt the following phenomenological expressions

\begin{equation}
 \label{G2}
 \Gamma_r= 3H \xi \exp[-(\alpha a^{n})],
 \end{equation}
and

\begin{equation}
 \label{G3}
 \Gamma_{dm} = 3H \beta [1 - \tanh(10-12a)],
\end{equation}
where, $\xi$, $\alpha$, $n$, and $\beta$ are positive constant parameters.
The parameter $n$ is associated with
the decay rate of the particle production. The higher $n$, the faster $\Gamma_r$ goes down.
Thus, in order to obtain an inflationary expansion for a sufficiently long period of time,
$n$ must lie in an interval $0.10 \leq n \leq 0.12$. Here,
we adopt $n = 0.10$. For other values within the above range,
the dynamics generated  by $\Gamma_r$ is practically the same.
Let us consider $\xi = 3/4$, to eliminate the possibility of an EoS less than $-1$ in the inflationary era.
The constants $\alpha$, $\beta$ are free parameters of the model to be constrained by the observational data.
When %the scale factor 
$a \rightarrow 0$, $\Gamma_{dm}$ gets close to zero regardless
the value of $\beta$. In fact, as we will see later, $\Gamma_{dm}$ practically does not influence the cosmic  dynamics
for $a \geq 0.6$, since in the very early universe, the production of non relativistic matter is negligible.
Figure \ref{Gamma_3H} shows the ratio $\Gamma_r/3H$ in terms of the scale factor for different values of $\alpha$.
From Eqs. (\ref{G}) -- (\ref{G3}), it is seen that at early times ($a \ll 1$) $\Gamma \simeq  \Gamma_r$, and
$\Gamma \simeq  \Gamma_{dm}$, otherwise.
\\

Let us assume that the dynamics of the very early universe was dominated by a production of particles given by Eq. (\ref{G2}).
At the end of this phase, the scale factor will have grown enormously, the production of particles will have declined sharply,
and the transition to the radiation dominated phase occurs. According to Parker's theorem, in the latter phase massless particles
cannot be quantum-mechanically produced \cite{Parker-Toms}.

The dynamics of the early inflationary era is usually described by a self-interacting scalar field, $\phi$, that slowly rolls down
its potential, $V(\phi)$, in such a way that the latter dominates the total energy density (i.e., $\dot{\phi^2}/2 \ll V(\phi)$)
during the inflationary expansion. It is therefore illustrative to relate $w_c$ to a scalar field that
would generate the same amount of inflation as the particle production scenario.

Since $\rho_{\phi} + p_{\phi} = \dot{\phi^2}$, and it must be equal to $\rho_{r} + p_{r}$, it follows that $\dot{\phi}^2=\rho_{r}(1+w_c)$,
where $\rho_{r}$ represents the energy density of the relativistic matter.

Hence,

\begin{equation}
\frac{d\phi}{da} = \frac{1}{a H} \sqrt{\rho_{r}(1+w_c)}.
\end{equation}

Accordingly,

\begin{equation}
\label{DF}
\Delta \phi = \int \frac{da}{a H} \sqrt{\rho_{r}(1+w_c)}.
\end{equation}

As a consequence, the total variation of the scalar field during inflation is

\begin{equation}
\label{delta_phi}
 \frac{\Delta \phi}{M_{pl}} \simeq \sqrt{3} \sqrt{\Big[ 1 + w_c \Big]} N ,
\end{equation}
where $N = \ln\Big(\frac{a_{end}}{a_i}\Big)$ is the total number %$N$ 
of e-folds produced during inflation.
Generically, total number of e-folds should be about 60 in order to solve the
flatness and horizon problem of the standard big bang theory. In fact, the spectrum of fluctuations observed in the CMB
corresponds to the values of $N$ in the interval $50 \leq N \leq 60$ \cite{Planck}.
\\

\begin{figure}
\centering
\includegraphics[width=3.2in, height=2.4in]{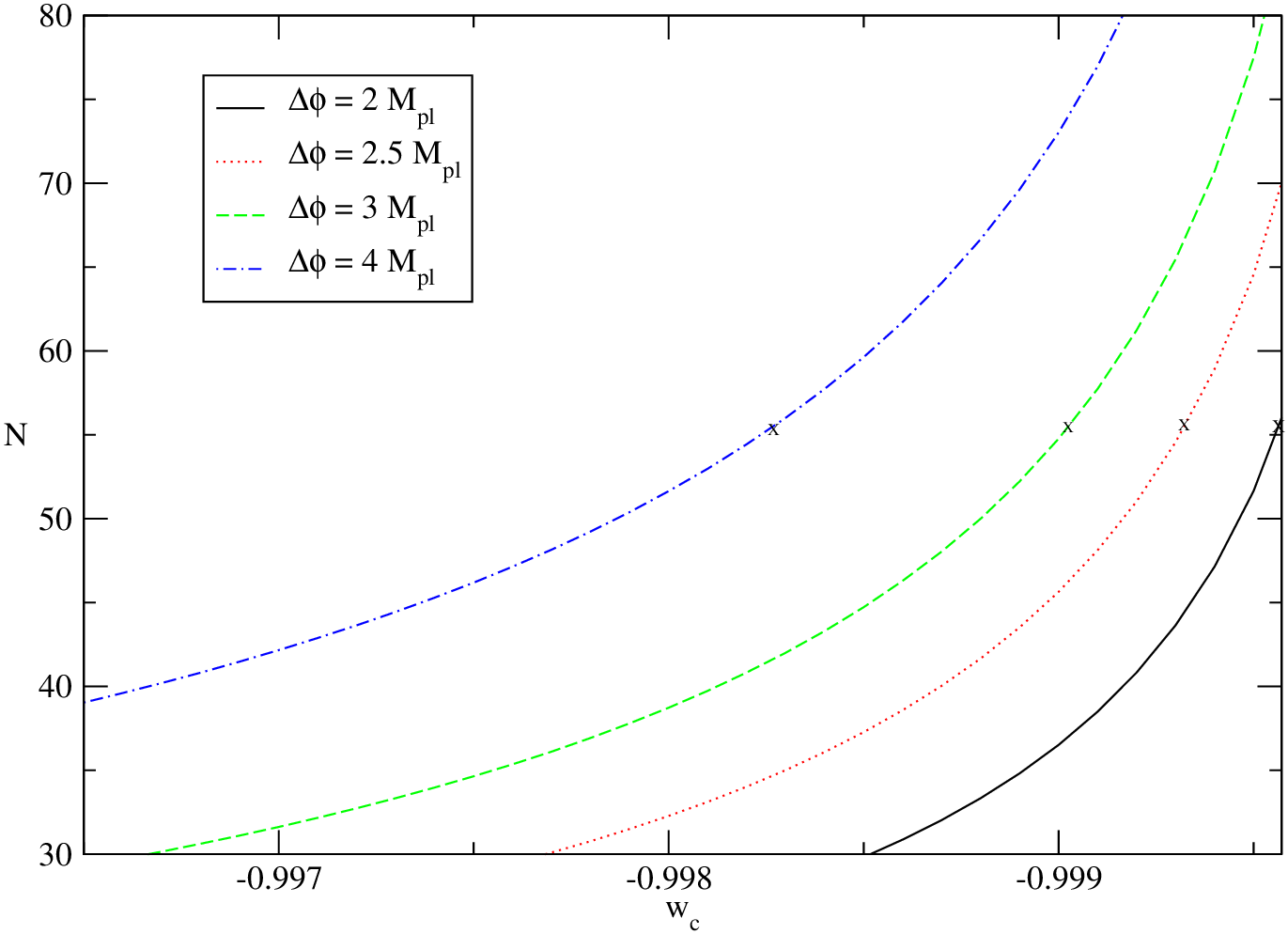}
\caption{\label{efols_CCDM_I}  Number of e-folds as a function of $w_c$.}
\end{figure}

Figure \ref{efols_CCDM_I} shows $N$ as a function of $w_c$ for different values of $\Delta \phi$.
Note that, to produce a suitable number of e-folds, we must have $w_c \simeq - 1$.
This is  possible provided, $0 < \alpha \leq 10$ (see Fig. \ref{Gamma_3H}). The number of e-folds
$N$ before the end of inflation is related to the variation of scalar field by

\begin{equation}
  dN = - H dt = \frac{1}{\sqrt{2 \epsilon} M_{pl}} d\phi,
\end{equation}
where $\epsilon = \frac{M^2_{pl}}{2}\Big(\frac{1}{V}\frac{dV}{d\phi}\Big)^2$ is a dimensionless slow-roll parameter.
The accelerated expansion occurs so long as $\epsilon \ll 1$. For later use, we recast the above equation as

\begin{equation}
 \label{epsilon1}
  \epsilon = \frac{1}{2} \Big[ \frac{d}{dN} \Big( \frac{\Delta \phi}{M_{pl}} \Big)  \Big]^2.
\end{equation}
\\

An obvious condition for the slow-roll is that $\ddot{\phi} << H \dot{\phi}$.
This requires that the so-called second slow-roll parameter, $\eta \equiv \epsilon - \frac{1}{H} \frac{1}{2 \epsilon} \frac{d \epsilon}{dt}$,
be much less than unity. This implies,

\begin{equation}
\label{eta}
\eta = \epsilon - \frac{1}{2 \epsilon} \frac{d \epsilon}{dN} \ll 1.
\end{equation}

During slow-roll regime, $\epsilon \ll 1$ and $\mid\eta\mid \ll 1$. From Eqs. (\ref{epsilon1}) and (\ref{delta_phi}), it follows

\begin{equation}
 \label{epsilon2}
  \epsilon = \frac{3}{2} \Big( 1 + w_c \Big),
\end{equation}
and $\eta = \epsilon$.

A further important parameter related to the inflationary behavior is the tensor-to-scalar ratio $r$,
which quantifies the ratio between the scalar and tensor spectra of the fluctuations produced in the inflationary era.
During the slow-roll, $r$ does not evolve much and one may recover Lyth bound \cite{Lyth} which relates $r$ to the
total field excursion during inflation

\begin{equation}
 \frac{\Delta \phi}{M_{pl}} \simeq \Big(\frac{r}{0.01}\Big)^{1/2}.
\end{equation}

From Eq. (\ref{delta_phi}), we have

\begin{equation}
 r = 3 \times 10^{-2} \, N^2 \, \Big(1 + w_c \Big).
\end{equation}

Considering $N =55$ (see Figure \ref{efols_CCDM_I}), we obtain $r$ = 0.163, 0.090, 0.062, and 0.045, respectively,
for $\Delta \phi = 4 M_{pl}$, $\Delta \phi = 3 M_{pl}$,
$\Delta \phi =2.5 M_{pl}$, and $\Delta \phi =2 M_{pl}$.
In the slow-roll approximation, keeping in mind that $\epsilon = \eta$ for the model presented in this work,
we have $r \simeq 8(1-n_s)$. Thus, we get $n_s \simeq 0.959, 0.977, 0.984$, and 0.988, for
$\Delta \phi = 4 M_{pl}$, $\Delta \phi = 3 M_{pl}$, $\Delta \phi =2.5 M_{pl}$, and $\Delta \phi =2 M_{pl}$, respectively.
The Planck collaboration obtained $r < 0.11$ (95 $\%$ CL), and $n_s = 0.968^{+0.006}_{-0.006}$ (68 $\%$ CL).
\\

The model presented in this section for the particle creation rate in the very early universe given
by Eq. (\ref{G2}) is in good agreement with the observational data recently released by the Planck collaboration \cite{Planck}.
The dynamical consequences of the second term in Eq.(\ref{G}), $\Gamma_{dm}$, that gives the production
rate of density matter particles, will be considered in section \ref{post_inflationary_era}.

\begin{figure}
\centering
\includegraphics[width=3.2in, height=2.4in]{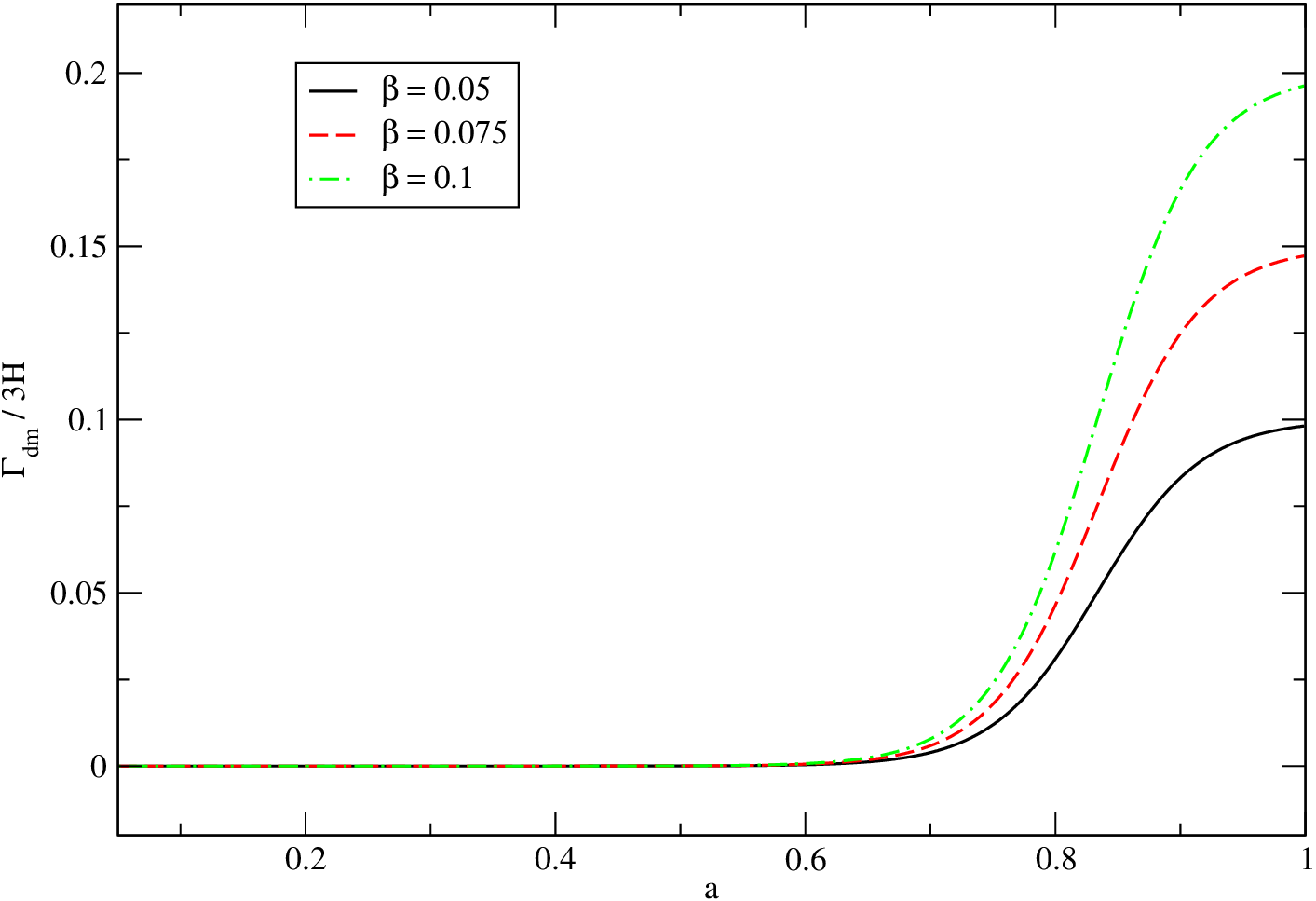}
\caption{\label{G_3H_dm} Evolution of ratio $\Gamma_{dm}/3H$.
In drawing the graphs, we have taken $\beta=0.050$, 0.075 and 0.1, for the solid, dashed, and dot-dashed line, respectively.}
\end{figure}

\section{The post inflationary era}
\label{post_inflationary_era}

Let us now consider a spatially flat FRW universe dominated
by pressureless matter (baryonic plus dark matter) and the energy
of the quantum vacuum (the latter with EoS $p_{\Lambda} = -
\rho_{\Lambda}$) in which a process of dark matter creation from
the gravitational field, governed  by
%%%%%%%%%%%%%%%%%%%%%%%%%%%%%%%%%%%%%%%%%%%%%%%%%%%%%%%%%%
\begin{equation}
\label{dark_matter_evolution} \dot{\rho}_{dm} \,  + \, 3H
\rho_{dm} = \rho_{dm} \, \Gamma_{dm},
\end{equation}
%%%%%%%%%%%%%%%%%%%%%%%%%%%%%%%%%%%%%%%%%%%%%%%%%%%%%%%%%%
is taking place. Since the production of ordinary particles is
much limited by the tight constraints imposed by local gravity measurements \cite{ellis, peebles_ratra, Hagiwara}, and
radiation has practically negligible impact on the recent cosmic dynamics, hence, for the sake of simplicity,
we assume that the produced particles are just dark matter particles. In writing the last equation, we used Eq.
(\ref{Eq:nbalance}) specialized to dark matter particles and the
fact $\rho_{dm} = n_{dm} \, m$, where $m$ stands for the rest mass
of a typical dark matter particle.
Figure \ref{G_3H_dm} shows the evolution of $\Gamma_{dm}/3H$ defined in Eq. (\ref{G3}) up to present moment ($a_0=1$)
for different values of $\beta$. %For example, we have that, at
In particular, for $\beta=0.1$ (shown in the dot-dashed line), we find that, 
at $a = 0.1$, $\Gamma_{dm}/3H \simeq 10^{-9}$; $a = 0.4$, $\Gamma_{dm}/3H \simeq 10^{-5}$; $a = 0.7$, $\Gamma_{dm}/3H \simeq 10^{-2}$; and finally, at
$a_0=1$, $\Gamma_{dm}/3H \simeq 0.1951$. 

%$a = 0.1$ $(\Gamma_{dm}/3H \simeq 10^{-9})$, $a = 0.4$ $(\Gamma_{dm}/3H \simeq 10^{-5})$, $a = 0.7$ $(\Gamma_{dm}/3H \simeq 10^{-2})$,
%and $a_0=1$ $(\Gamma_{dm}/3H \simeq 0.1951)$, for $\beta=0.1$ shown in the dot-dashed line.
Since baryons are neither created nor destroyed, their corresponding energy density obeys
$\dot{\rho}_{b} \, + \, 3H \rho_{b} = 0$. On the other hand,  as the
energy of the vacuum does not vary with expansion, so $\, \rho_{\Lambda} = {\rm constant}$.
\\

In this scenario the total pressure is $\, p_{\Lambda}
\, + \, p_{c}$, thereby the effective EoS is just the sum of the
EoS of vacuum plus that due to the creation pressure,
%%%%%%%%%%%%%%%%%%%%%%%%%%%%%%%%%%%%%%%%%%%%%%%%%%%%%%%%%%%%%%%%%%%%%%
\begin{equation}
w_{eff} = -1 \, - \frac{\Gamma}{3H} \, .
\label{Eq:weff}
\end{equation}
%%%%%%%%%%%%%%%%%%%%%%%%%%%%%%%%%%%%%%%%%%%%%%%%%%%%%%%%%%%%%%%%%%%%%%%
Since, by the second law of thermodynamics, $\Gamma$ is positive-semidefinite, we
have that the effective EoS can be less than $-1$ without the need
of invoking any scalar field with wrong sign in the kinetic
term. Therefore, due to the combined effects of the vacuum and
creation pressures, one can hope for a global EoS less that $-1$ without the need of any phantom fields.
We mention that, an equation of state $ w_{de} < -1 $, without any introduction of phantom fields was first realized in 
the context of modified gravity models \cite{Boisseau}, more specifically in the scalar-tensor theory of gravity.
However, it is worth to recall that, %it is important to mention that, 
we obtain this phantom behavior without %the need for any modification of gravity theory,
any modifications in the gravitational theories, rather by the mechanism of gravitational particle productions 
in an adiabatic manner.
%where $w_{de} < -1$ and simply obtained by the particle production process. 
We note that, particle creation in an expanding universe has been discussed to understand several aspects 
of modern cosmology.
Recently, the authors in Refs. \cite{Capozziello,Pereira,Singh} investigated the particle production rate in the context of
$f(R)$ gravity. On the other hand, the possible effects of this mechanism have also been analyzed in case of a 
flat and negative curved FRW universes as well \cite{Montani}.
\\

Friedmann's equation for this scenario is,

\begin{equation}
\label{fields_Einstein1}
 H^2=\frac{8 \pi G}{3} (\rho_{b} \, + \, \rho_{dm} \, + \,
 \rho_{\Lambda}).
\end{equation}

Inserting (\ref{G3}) in (\ref{dark_matter_evolution}), and integrating, we have

\begin{equation}
\label{dark_matter_evolution_2} \rho_{dm}=\rho_{dm0}\, a^{-3} \,
\exp \Big(3 \beta \int_{1}^{a} \frac{k(\tilde{a})}{\tilde{a}} d
\tilde{a} \Big),
\end{equation}
%%%%%%%%%%%%%%%%%%%%%%%%%%%%%%%%%%%%%%%%%%%%%%%%%%%%%%%%%%%%%%%%%%%%%%%%%%%%%%%%%%%%%%%%%%%%%
where $k(a) = \Gamma_{dm}/3H = \beta[ 1 - \tanh(10-12a)]$.

In terms of the redshift, $z=a^{-1}-1$, the Hubble
expansion rate reads

\begin{eqnarray}
\label{Hz_model_2}
 \frac{H^2(z)}{H^2_0}=\Omega_{b0} \, (1+z)^3 \nonumber\\+  \, \Omega_{dm0}\, (1+z)^3 \,
 \exp \Big(-3 \beta \int_{0}^{z} \frac{k(\tilde{z})}{(1+\tilde{z})} d\tilde{z} \Big)\,
 + \, \Omega_{\Lambda0},
\end{eqnarray}
where the $\Omega_{i0}$ denote the current
fractional densities of baryons, dark matter and vacuum, respectively.

\section{DATA SETS AND STATISTICAL ANALYSES}
\label{Statistical_Analysis}

We first describe the set of data used in the statistical analysis. To constrain the free parameters
$\theta_i=\{\beta,\Omega_{dm0} \}$ of the particle creation model, we use the following sets of data:
a) The most recent Type Ia Supernovae (SNe Ia) data sets from the joint light-curve analysis (JLA) \cite{snia};
b) Baryon acoustic oscillations (BAO) data from the SDSS Luminous Red Galaxy sample \cite{Blake}, the
WiggleZ Survey \cite{Percival} and 6dF Galaxy Survey \cite{Beutler};
c) Measurements of the Hubble function $H(z)$ compiled in \cite{Hz}, plus data by the
BOSS collaboration \cite{BOSS}, $H(z=2.34)= 222 \pm 7$ km $s^{-1}$ Mpc. Their best fit values, with their
corresponding 1$\sigma$ uncertainties are presented in subsection \ref{results}.
These follow from minimizing the likelihood function
$L \propto \exp(-\chi_{total}^2)$ with $\chi_{total}^2= \chi_{SNIa}^2+\chi_{BAO/CMB}^2+\chi_{H}^2$,
where each $\chi^2_i$ (specified below) quantifies the discrepancy between theory and observation.
The statistical analysis used for those observables is described in the following subsections.

\subsection{Supernovae type Ia}

SNe Ia are very bright standard candles, useful for measuring cosmological distances.
Here, we use the JLA compilation consisting of 740 well-calibrated SNe Ia in the redshift range $z \in [0.01, 1.30 ]$ \cite{snia}.
This collection of SNe Ia includes about 100 low-redshift SNe from a combination of various subsamples,  $\sim$ 350
from SDSS at low to intermediate redshifts,  $\sim$ 250 from SNLS at intermediate to high redshifts, and
$\sim$ 10 high-redshift SNe from the Hubble Space Telescope. All of the SNe Ia have light curves of high quality,
so their distance moduli can be obtained accurately. These data points are the most recent SNe Ia present in the literature.

The distance modulus predicted for a given supernova of redshift $z$ can be expressed as

\begin{equation}
\mu(z, \theta_i)= 5 \log_{10} \Big[ \frac{d_L(z,\theta_i)}{Mpc} \Big] + 25
\end{equation}
where $d_{L} = (1+z)H_0\int_0^z \frac{dz'}{H(z')}$ is the luminosity distance.

The corresponding $\chi^2$ is then calculated in the usual way for correlated observations:

\begin{equation}
\chi^2_{SNIa}= \Delta \mu^{\dag} C^{-1} \Delta \mu,
\end{equation}
where $\Delta \mu = \mu_{obs} - \mu_{th}(\theta_i)$ is the vector of differences between the observed, corrected distance
moduli and the theoretical predictions that depend on the set of cosmological model parameters $\theta_i$, and $C$ is
the covariance matrix for the observed distance moduli. The latter can be found in \cite{snia}.

\subsection{Baryon acoustic oscillations and cosmic microwave background}

Here, we use a more model-independent constraint derived from the product of the acoustic scale of the
cosmic microwave background (CMB), $l_A = \pi d_A (z_{*})/r_s (z_{*})$, and the measurement of the ratio of the sound horizon
scale at the drag epoch and the BAO dilation scale, $r_s (z_d)/D_V(z_{BAO})$,
defined as $X=\frac{ d_A (z_{*})}{D_V(z_{BAO})}\frac{r_s (z_d)}{r_s (z_{*})}$. Here, $D_V(z)=[d^2_A(z)cz/H(z)]^{1/3}$
is the dilation scale introduced in \cite{Daniel}, $d_A(z_{*})$ is the comoving angular-diameter distance to recombination
$d_A(z_{_{*}})=c \int_0^{z_{*}} \frac{dz'}{H(z')}$, and $r_s(z_{*})$ is the comoving sound horizon at decoupling

% \begin{equation}
% d_A(z_{_{*}})=c \int_0^{z_{*}} \frac{dz'}{H(z')},
% \end{equation}
% and $r_s(z_{*})$ is the comoving sound horizon at decoupling

\begin{equation}
r_s(z_{*})= \frac{c}{\sqrt{3}} \int_0^{1/(1+z*)} \frac{da}{a^2H(a)\sqrt{1+(3\Omega_{b0}/4\Omega_{\gamma0})}}.
\end{equation}

Inserting the ratio $r_s(z_d)/r_s (z_{*}) = 1.044 \pm 0.019$, with $z_d= 1020$ and $z_{*}=1091$ \cite{Bennett},
in the above equation for $X$, we obtain the BAO/CMB constraints

\begin{equation}
X=\frac{d_A(z_{*})}{D_V(z_{BAO})}.
\end{equation}

We write the $\chi^2$ for the BAO/CMB analysis as

\begin{equation}
\chi^2_{BAO/CMB}= \Delta X^{\dag} C^{-1} \Delta X,
\end{equation}
where $\Delta X = X_{obs} - X_{th}(\theta_i)$, and $C^{-1}$ is the inverse covariance matrix obtained in \cite{Giostri}. Here we make use of the six data point compiled by R. Giostri {\it et al}. \cite{Giostri}.

% \[ \left( \begin{array}{cccccc}
% 0.48435    & -0.101383  & -0.164945   & -0.0305703 & -0.097874 & −0.106738 \\
% -0.101383  & 3.2882     & -2.45497    & -0.0787898 & -0.252254 & -0.2751 \\
% -0.164945  & -2.45497   & 9.55916     & -0.128187  & -0.410404 & -0.447574 \\
% -0.0305703 & -0.0787898 & -0.128187   & 2.78728    & -2.75632  & 1.16437 \\
% −0.097874  & −0.252254  &  −0.410404  & −2.75632   & 14.9245   & −7.32441 \\
% −0.106738  & −0.2751    &  −0.447574  &  1.16437   & −7.32441  & 14.5022  \end{array} \right)\]
% is the inverse covariance matrix obtained in \cite{Giostri}. Here we make use of the six data point compiled by R. Giostri {\it et al}. \cite{Giostri}.

\subsection{History of the Hubble parameter}

The differential evolution of early-type passive
galaxies provides direct measurements of the Hubble parameter, $H
(z)$. An updated compilation of such data was presented in
\cite{Hz}. We adopt 28 data points in the redshift range
$0.09 < z < 1.75$ reported in \cite{Hz}, plus data from the
BOSS colaboration \cite{BOSS}, $H(z=2.34)= 222 \pm 7$ km $s^{-1}$ Mpc, to constrain the free parameters
$\theta_i$. We compute the $\chi^2_H$ function defined as

\begin{equation}
\label{chi_quadrado_H}
\chi^2_{H}(\theta_{i},H_{0})=\sum_{i=1}^{28}
\frac{[H^{obs}(z_i)-H^{th}(z_i,H_0,\theta_{i})]^2}{\sigma^2_{H}(z_i)},
\end{equation}
where $H_{th} (z_i , H_0 , \theta_i)$ is the model-predicted value of the Hubble parameter at the redshift $z_i$.
The present value of the Hubble parameter $H_0$ is marginalized.

\subsection{Statistical results}
\label{results}
\begin{figure}
\centering
\includegraphics[width=3.2in, height=2.4in]{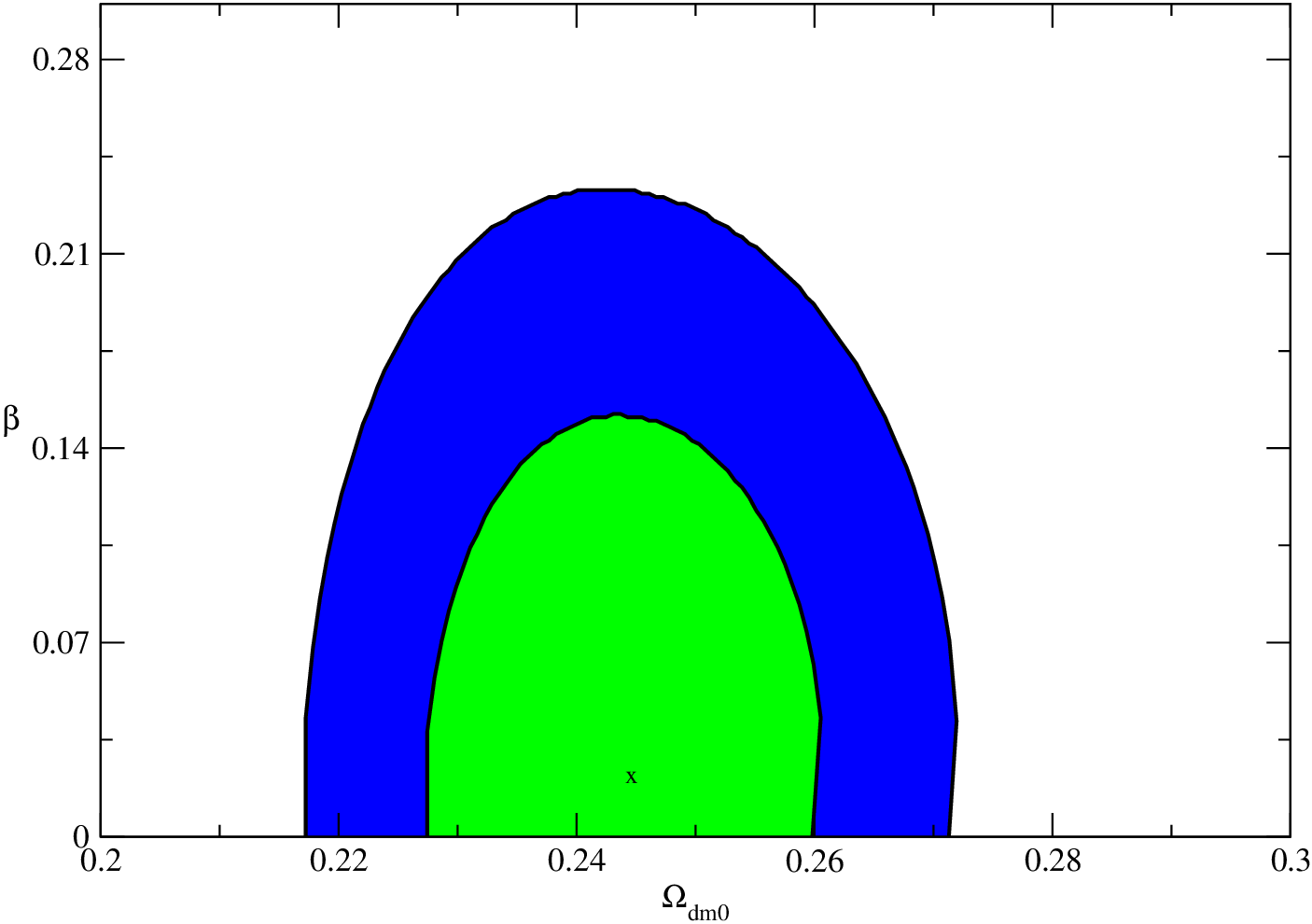}
\caption{\label{constrain1} 1$\sigma$ and 2$\sigma$ confidence regions from joint analysis SNIa+BAO/CMB+$H(z)$.}
\end{figure}

\begin{figure}
\centering
\includegraphics[width=3.2in, height=2.4in]{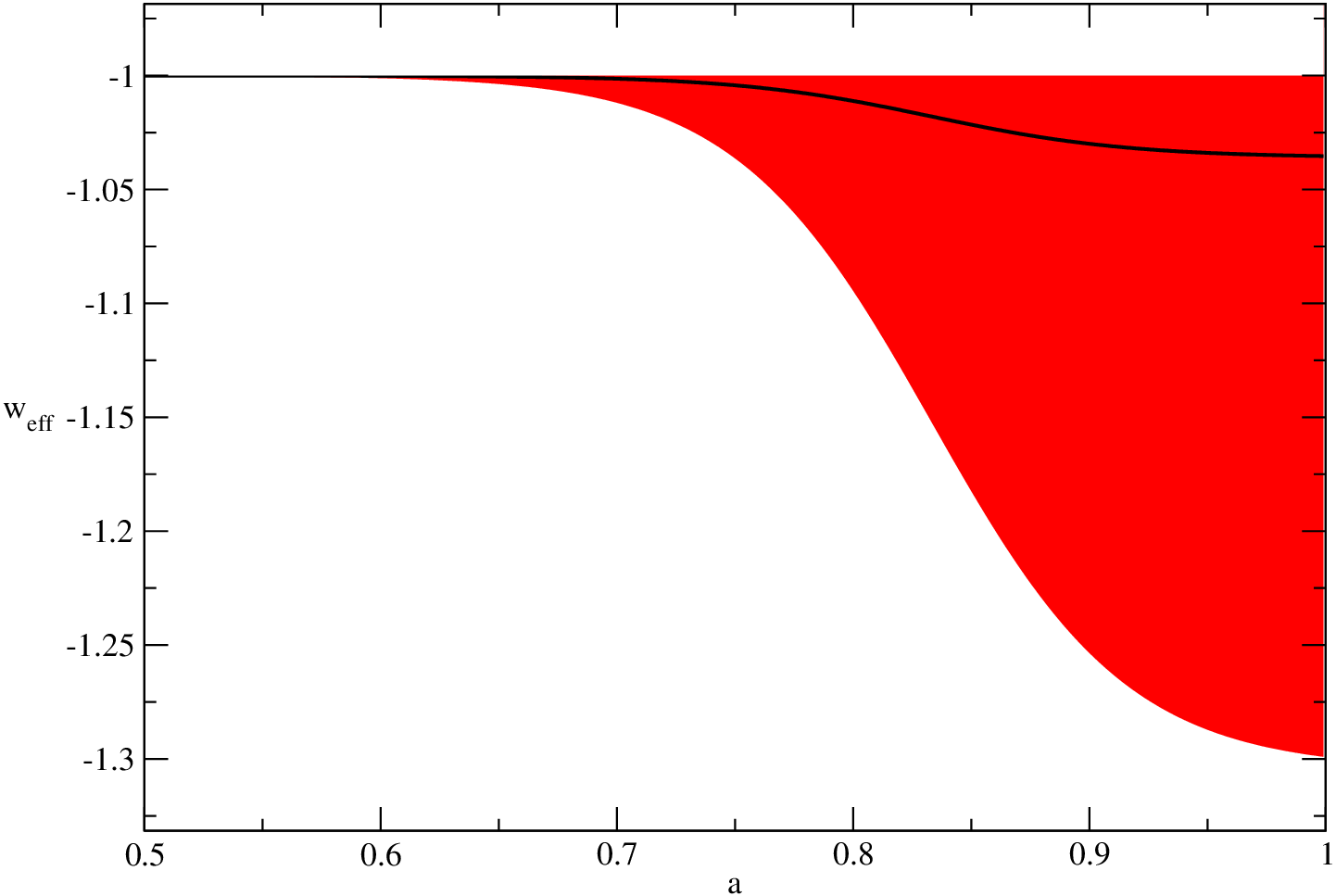}
\caption{\label{constrain2} Evolution of the effective EoS; the solid (black) line indicates the best fit value,
and the shaded region (red) the 1$\sigma$ uncertainty.}
\end{figure}

Figure \ref{constrain1} shows the 68 $\%$ and 95$\%$ confidence contours in the $\Omega_{dm0}-\beta$ plane.
During the statistical analysis is taken with physical condition which $\beta \geq 0$.
We obtain as best fit for the model, $\beta=0.018^{+0.134}_{-0.151}$ and $\Omega_{dm0}=0.242^{+0.018}_{-0.014}$ at 1$\sigma$
confidence level (CL). % with $\chi^2_{min}/dof=0.872$.
Although the best fit for $\beta$ is small, the possibility of a small creation rate ($\beta > 0 \rightarrow \Gamma > 0$)
not can be ruled out. In fact, $0 < \beta < 0.1523$ and $0 < \beta < 0.2328$ at 1$\sigma$ and 2$\sigma$ CL, respectively.
Figure \ref{constrain2} shows the reconstruction of the effective EoS  in terms of scale factor. Note that
$w_{eff}(a=1)=-1.053^{+0.053}_{-0.397}$, i.e, $w_{eff} < -1$ without the need  of invoking phantom fields.

% \begin{figure} % figuur 1
% \begin{minipage}{\columnwidth}
% \centering
% \framebox[\columnwidth][c]{\raisebox{0pt}[20mm][20mm]{ICCDM.eps}}
% \end{minipage}
% \caption{ 1$\sigma$ and 2$\sigma$ confidence regions from joint analysis SNIa+BAO/CMB+$H(z)$.}
% \label{constrain1}
% \end{figure}
%
% \begin{figure} % figuur 1
% \begin{minipage}{\columnwidth}
% \centering
% \framebox[\columnwidth][c]{\raisebox{0pt}[20mm][20mm]{w_eff_I.eps}}
% \end{minipage}
% \caption{ Evolution of the effective EoS; the solid (black) line indicates the best fit value, and the shaded region (red) the 1$\sigma$ uncertainty.}
% \label{constrain2}
% \end{figure}

\section{Conclusions}

As shown by Parker and collaborators, the particle creation is something expected in expanding spacetime \cite{Parker}.
In spite of the difficulty in deriving the production rate, this phenomenon may in principle be related with inflation and
the present cosmic acceleration \cite{ccdm_model}.
\\

However, the particle creation processes, 
during the very rapid early expansion of the universe, are believed to give rise to temperature
anisotropies in the cosmic microwave background. Within of the context of a continuous matter creation process
in an expanding universe, the effects on the CMB TT and EE power spectra were first investigated recently in \cite{eu_pan}.
In the cosmological context, CMB can be a powerful source to investigate the properties of an adiabatic matter 
creation process to strengthen the cosmological models driven by the adiabatic particle productions both 
for early and late universe.
\\

We have shown %in this paper 
here that in the limit of high energies, the production of relativistic particles from
the vacuum leads to a viable inflationary solution, and this dynamics is in good agreement with the observational
data recently released by the Planck collaboration \cite{Planck}.
Further, we present an alternative to the recently reported values of 
the dark energy equation of state beyond $-1$, 
%that lower than $-1$ for the equation of state of dark energy
may arise from the  joint effect of the quantum vacuum and the process of  particle production.
This offers a viable alternative to the  embarrassing possibility of the scalar fields %that, among other things,
which violate the dominant energy condition, and give rise to classical and quantum instabilities, 
and further do not respect the second law of thermodynamics.
\\

%To sum up, 
Summarizing, by proper choice of the particle creation rate, the cosmic scenario  presented in this work shows the evolution
of the universe starting from the early inflationary era to the present accelerating phase, considering a continuous matter creation
process by the gravitational field. Obviously, phenomenological models of particle production different from the ones essayed
here are also worth exploring. %More important, however, is to determine the rate $\Gamma$ using quantum field theory.
However, the most important thing, from which the cosmological scenarios could be viewed more clearly,
is to determine the production rate $\Gamma$ using quantum field theory in curved spacetimes.
\\
% , but, as said above,
% this does not seem feasible until the nature of  dark matter particles is found.

\begin{acknowledgements}
 I am grateful to the financial support from CAPES Foundation Grant No. 13222/13-9 and from the Dept. ECM, Universitat de Barcelona.
 The author is very grateful to D. Pav\'on for his useful comments and S. Pan for a critical reading of the
 manuscript. Finally, the author is also very grateful to the anonymous referee for several comments
 which improved the manuscript considerably.
\end{acknowledgements}

\end{document}